\documentclass[conference]{IEEEtran}
\IEEEoverridecommandlockouts
\usepackage{cite}
\usepackage{amsmath,amssymb,amsfonts}
\usepackage{algorithmic}
\usepackage{graphicx}
\usepackage{textcomp}
\usepackage{xcolor}
\usepackage{soul}
\usepackage{caption}
\usepackage{subcaption}
\usepackage{tabu} 
\usepackage{booktabs} 
\def\BibTeX{{\rm B\kern-.05em{\sc i\kern-.025em b}\kern-.08em
    T\kern-.1667em\lower.7ex\hbox{E}\kern-.125emX}}
\begin{document}

\title{Evaluating Deep Networks for Detecting User Familiarity with VR from Hand Interactions}

\author{
    \IEEEauthorblockN{Mingjun Li, Numan Zafar, Natasha Kholgade Banerjee, Sean Banerjee}
    \IEEEauthorblockA{Clarkson University, Potsdam, NY, USA
    \\\texttt{\{mingli,zafarn,nbanerje,sbanerje\}@clarkson.edu}}
}

\maketitle

\begin{abstract}
As VR devices become more prevalent in the consumer space, VR applications are likely to be increasingly used by users unfamiliar with VR. Detecting the familiarity level of a user with VR as an interaction medium provides the potential of providing on-demand training for acclimatization and prevents the user from being burdened by the VR environment in accomplishing their tasks. In this work, we present preliminary results of using deep classifiers to conduct automatic detection of familiarity with VR by using hand tracking of the user as they interact with a numeric passcode entry panel to unlock a VR door. We use a VR door as we envision it to the first point of entry to collaborative virtual spaces, such as meeting rooms, offices, or clinics. Users who are unfamiliar with VR will have used their hands to open doors with passcode entry panels in the real world. Thus, while the user may not be familiar with VR, they would be familiar with the task of opening the door. Using a pilot dataset consisting of 7 users familiar with VR, and 7 not familiar with VR, we acquire highest accuracy of 88.03\% when 6 test users, 3 familiar and 3 not familiar, are evaluated with classifiers trained using data from the remaining 8 users. Our results indicate potential for using user movement data to detect familiarity for the simple yet important task of secure passcode-based access.
\end{abstract}


\section{Introduction}

Virtual reality (VR) is increasingly being looked at as a mechanism for delivering experiences for tasks that users may typically perform in real-world, desktop, and mobile environments. VR environments are being evaluated across a diverse spectrum of users for education~\cite{rojas2023systematic}, therapy~\cite{chard2022virtual,liu2022virtual}, physical fitness~\cite{ng2019effectiveness}, and even applications such as security~\cite{jones2021literature,stephenson2022sok} and personal banking~\cite{mathis2022can}. Research in behavior-based VR security shows that user actions in VR change at varying timescales~\cite{liebers2023exploring,miller2022temporal,miller2023large}. As VR applications become more prevalent in consumer spaces, the level of first-time or early-stage VR users is expected to rise. For instance, an older adult may be recommended an exercise routine they can perform in a VR space. If the user is unfamiliar with the VR application, they may need timely intervention to provide training as otherwise they may become disincentivized and stop using the application. Though such training can be given by the service provider, automatic training delivery by detecting the familiarity of the user provides the benefit of providing on-demand training in the user's personal environment, and alleviates the burden on strained service staff. 

In this work, we provide a first attempt at conducting automated familiarity detection using the movements of a user as they engage in a VR task. We use a door opening VR task, where a user enters a 4-digit passcode combination to open a VR door, as users, regardless of prior VR experience, will have performed similar tasks in the real world. We envision VR doors to be a point of entry to collaborative virtual spaces. For example, a user in a virtual office setting may `walk' to a common virtual conference room and enter a combination before entering the room. Early detection of prior VR experience, in this case through the interaction with the VR door, could enable real-time modifications to the interaction elements before the user enters the conference room to perform more complex interactions. In our approach, we track the finger movements of a user entering a passcode combination to unlock a VR door, and train deep neural networks to detect familiarity with VR. We use user-reported binary experience level with VR as a representation of familiarity. The task considered, i.e., unlocking a VR door, also has special significance in the area of VR security. Recognizing the potential for VR applications to store sensitive user data, a number of VR applications investigate leveraging the VR environment to enable secure access, through passcodes or biometric signatures~\cite{jones2021literature,stephenson2022sok}. With the emergence of hand tracking for VR, recent work has explored user identification based on tracking hand data~\cite{liebers2022identifying} In trying to gain access to a secured space in VR, a novice user may perform actions that cause them to be locked out by VR authentication mechanisms, e.g., pressing the wrong keypad button due to lack of knowledge on interaction process, or performing a deviating movement. %Our work investigates the potential for movement-based familiarity detection which can be used to ensure that genuine users who may not be familiar with the VR environment are not locked out of secure spaces. 

To the best of our knowledge, no work exists for learning-based detection of user familiarity with VR. Some work exists in an allied area of using machine learning for familiarity assessment for a particular task simulated in a VR environment, e.g., evaluating surgical skill. For instance, a considerable body of work exists to detect surgical skill, by tracking the eye or hand movements of the practitioner~\cite{chan2021systematic}. Work similarly exists in using eye movements to detect soccer player expertise in VR~\cite{hosp2020eye,hosp2020eye2}. Work has evaluated assessing spatial familiarity during wayfinding based on the eye movements performed by a user when turning at junctions~\cite{alinaghi2022can}. However, the focus is on familiarity with the task, rather than with VR as an interaction medium itself. Work on understanding how different gaming controllers influence perceptions of usability have been explored~\cite{gerling2011measuring} and provide insights on how changes in ergonomics can impact usage.

\section{VR Door Unlock Application}
\begin{figure}
    \centering
        \includegraphics[width=\linewidth]{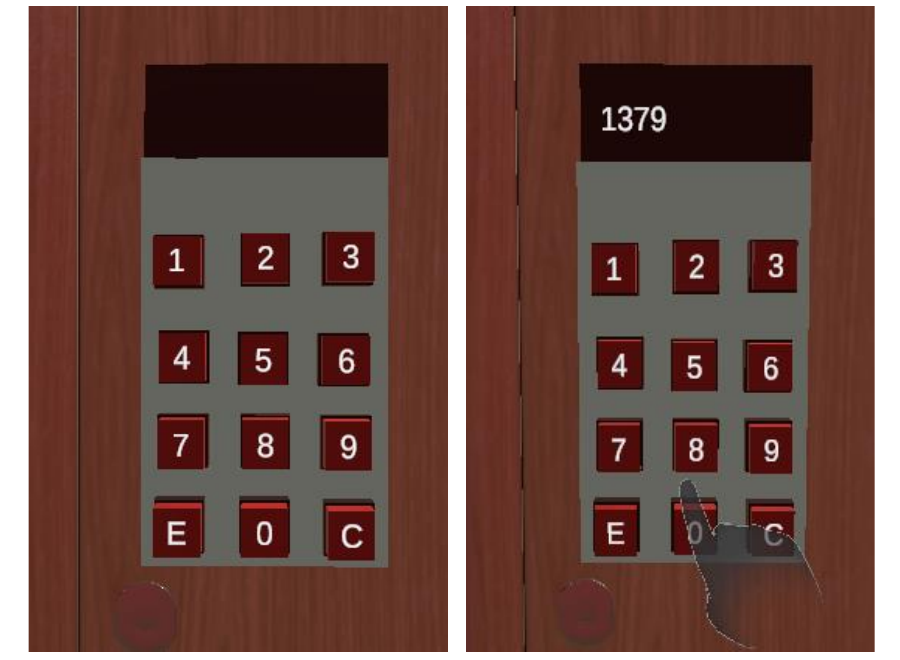}
        \caption{Left: Close-up view of numeric panel in VR Door Unlock application. The panel is placed on a VR door. Right: Participant interacts with the panel by entering a numeric passcode to unlock the door.}
        \label{fig:lock}
\end{figure}

We create a VR environment where users interact with a numeric panel to unlock a VR door via hand tracking using a controller-free head-mounted device (HMD). We use the Meta Quest Pro in this study. As VR environments become pervasive, an access controlled door may be used to allow users with varying familiarity in VR to enter a virtual meeting room or their office. Users who are unfamiliar with VR would still be familiar with the task from the real world as most people have opened a door using a numeric panel. To unlock the door, the user uses their (controller-free) hands to enter a numeric passcode combination by interacting with VR buttons simulating those on a traditional access panel on a real-world door. Upon completing the entry, they press a key labeled `E' to enter the combination and unlock the door. If the user enters the correct combination, the door opens immediately, and automatically closes and locks after 3 seconds. We design our panel's buttons to consist of block game objects  for digits 1 through 0, and the letters `E' represent enter and `C' representing clear. We use native hand tracking enabling the participant to use hand motions to interact with the game objects. We detect if a user has interacted with a button by detecting collisions using the collider on the tip of the index finger on both hands and the colliders on the buttons. Figure~\ref{fig:lock} shows a view of the door unlock application. 

\section{Dataset}

\begin{figure*}
     \centering
     \includegraphics[width=\textwidth]{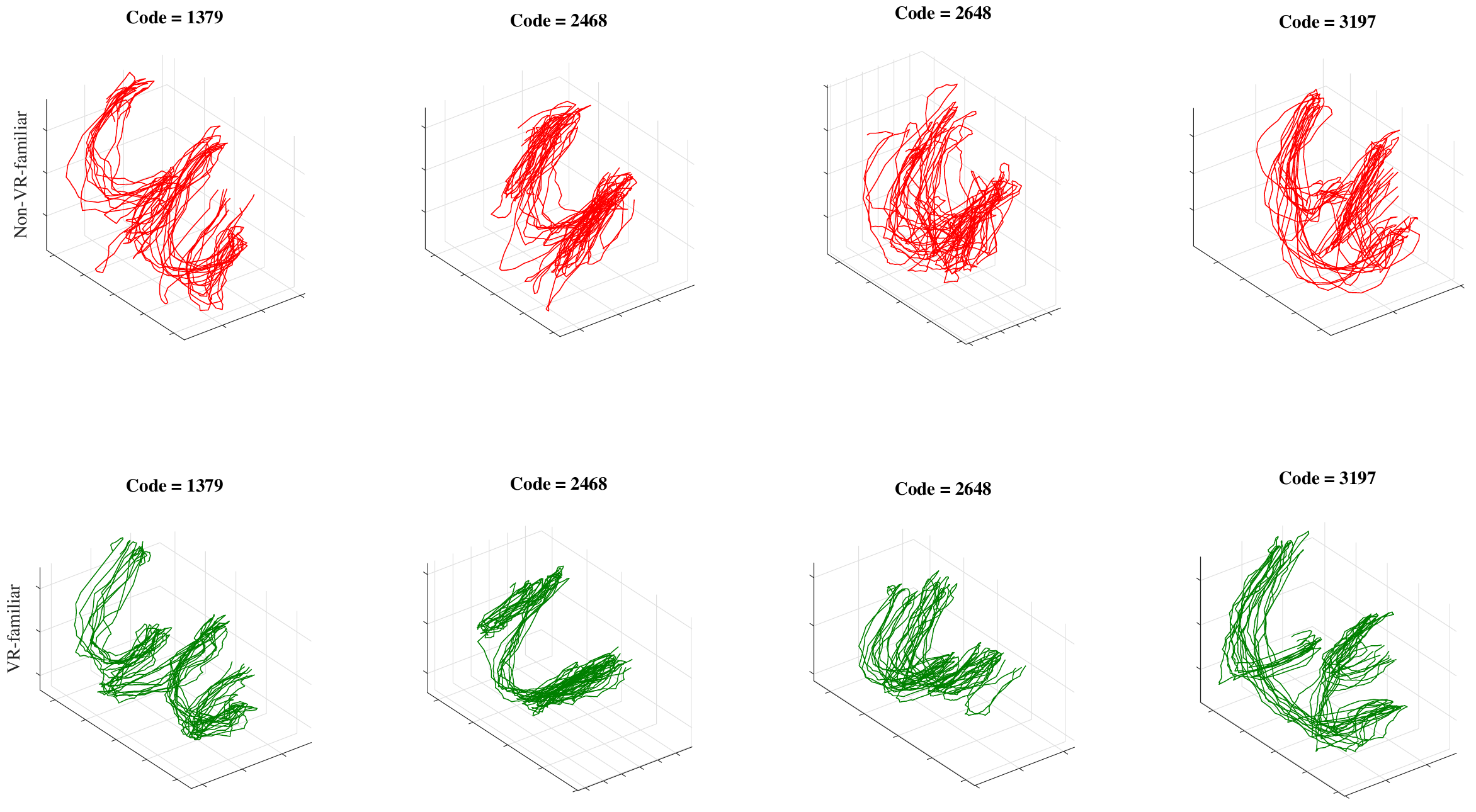}
     \caption{10-session trajectories from each of the 4 passcode combinations used in this work from a participant without VR familiarity (top) and with VR familiarity (bottom). 
     We observe that the user with no prior familiarity with VR show more variability in their interaction profile when compared to the user with prior familiarity.}
     \label{fig:trajs}
\end{figure*}

We recruited 14 participants from the student, faculty, and staff at Clarkson University. Prior to using the VR environment, we asked participants to indicate if they had prior experience in VR. Of the 14 participants, 7 reported no prior experience in VR, i.e., no familiarity, and the remaining 7 indicated being very familiar with VR. All participants in our study are right-handed with 10 participants self reporting as male and the remaining 4 as female. We collected data using a Meta Quest Pro and recorded position and orientation data at 60 frames per second. During data collection, we asked participants to enter $2648$, $2468$, $1379$, and $3179$ on the virtual keypad. Each combination was entered 10 times before moving to the next combination. Once the participant had entered a combination, they were asked to press `E' to open the door. If the participant made a mistake, we asked them to press `C' to clear the code and try again. Incorrect entries were stored, but not used during the training and testing phase. In Figure~\ref{fig:trajs} we show all 10 trials for each door lock combination for one VR-familiar and one non-VR-familiar participant. We observe that the user who self reported as unfamiliar with VR shows a more variable pattern of movement. 

\section{Experiments}
We evaluate detection of VR familiarity by training  classifiers on sliding windows extracted from the trajectory of the dominant hand of the participants in the dataset. We evaluate sliding windows of sizes 50, 60, 70, 80, 90, 100, 110, and 120, with a step size of 1. For each sliding window choice, we evaluate three types of classifiers\textemdash{}multi-layer perceptrons (MLPs), fully convolutional networks (FCNs)~\cite{wang2017time}, and Point Cloud Transformer (PCT)~\cite{guo2021pct}. We train one classifier per sliding window and per key combination.

%We partition the trajectories of the right hand of all participants in the dataset, which can also be considered as spatial and temporal data, by employing a sliding window of dimensions $ws*n_{feat}$, where $ws$ represents the length of timestamps of the trajectory and $n_{feat}$ denotes the number of features. This window adeptly maneuvers with a step size of $s$, designed to traverse along the time axis. We set the step size at $1$ across all experiments. Our intention is to rigorously investigate the potential impact of trajectory length variations on the resulting classification outcomes. We consistently apply the same set of values for the window sizes for all our experimental approaches, that is $ws = \{50, 60, 70, 80, 90, 100, 110, 120\}$. In the process of training the models, for every distinct password combination used by all users, we assign the first $5$ sessions to constitute the training set, while the subsequent $5$ sessions are earmarked for the testing set. 

%In this study, we employ a set of well-established classification models known for their robust performance and accuracy. These models include Multi-Layer Perceptron (MLP)~\cite{haykin1998neural}, Fully Convolutional Networks (FCNs)~\cite{wang2017time}, and Point Cloud Transformer (PCT)~\cite{guo2021pct}. 

\paragraph{Multi-Layer Perceptron}We design our MLP to consist of $2$ position-wise dense feed-forward layers where the dimension of the output for each layer is set at half of the input dimension. We retain this pattern across the hidden layers. Following each hidden layer, we apply a ReLU~\cite{agarap2018deep} activation layer. The output layer followed by a softmax layer to obtain the predicted class probabilities.

\paragraph{Fully Convolutional Network}We employ the FCN architecture as described by Wang et al.~\cite{wang2017time}. The architecture consists of $3$ convolutional blocks, each featuring a convolutional layer with a filter size of \{128, 256, 128\} and a 1D kernel with the size of \{8, 5, 3\}. We apply batch normalization layers~\cite{ioffe2015batch} after each convolutional layer. We use ReLU activation layer at the end of each block. After the $3$ blocks, we use a global average pooling layer~\cite{lin2013network} and a softmax operation for the final class probabilities.

\paragraph{Point Cloud Transformer}The Point Cloud Transformer (PCT)~\cite{guo2021pct}, created to address point cloud inputs via an attention architecture, consists of 4 stacked attention blocks. The outputs of each attention block are concatenated, additional max-pooling and average-pooling operations are applied, and the output is passed through several linear layers to produce the final classification probabilities. Given that our dataset contains trajectories that occupy a constrained space, we reduce the complexity of the PCT model by removing the concatenation of attention module outputs and eliminating the subsequent max-pooling and average-pooling layers.

\paragraph{Training and Test Split}

To ensure independence of users between the training and test sets, we randomly select a subset of 4 VR-familiar users and 4 non-VR-familiar users for training, and use the remaining 3 VR-familiar users and 3 non-VR-familiar users for testing. For each user, we use all 10 interaction sessions. 

\paragraph{Loss Function}
We optimize the parameters of each model by minimizing the Binary Cross-Entropy (BCE) loss,
\begin{equation}
     \mathtt{L} = (1/|W|) \Sigma_{W} BCE(pred, gt),
     \label{eq:bceloss}
\end{equation}
where $|W|$ denotes the number of windows, $pred$ is the predicted label, and $gt$ represents the ground truth label. We use Adam~\cite{kingma2014adam} to optimize the parameters of all the models.

\section{Results}

\begin{figure*}
    \centering
    \includegraphics[width=\linewidth]{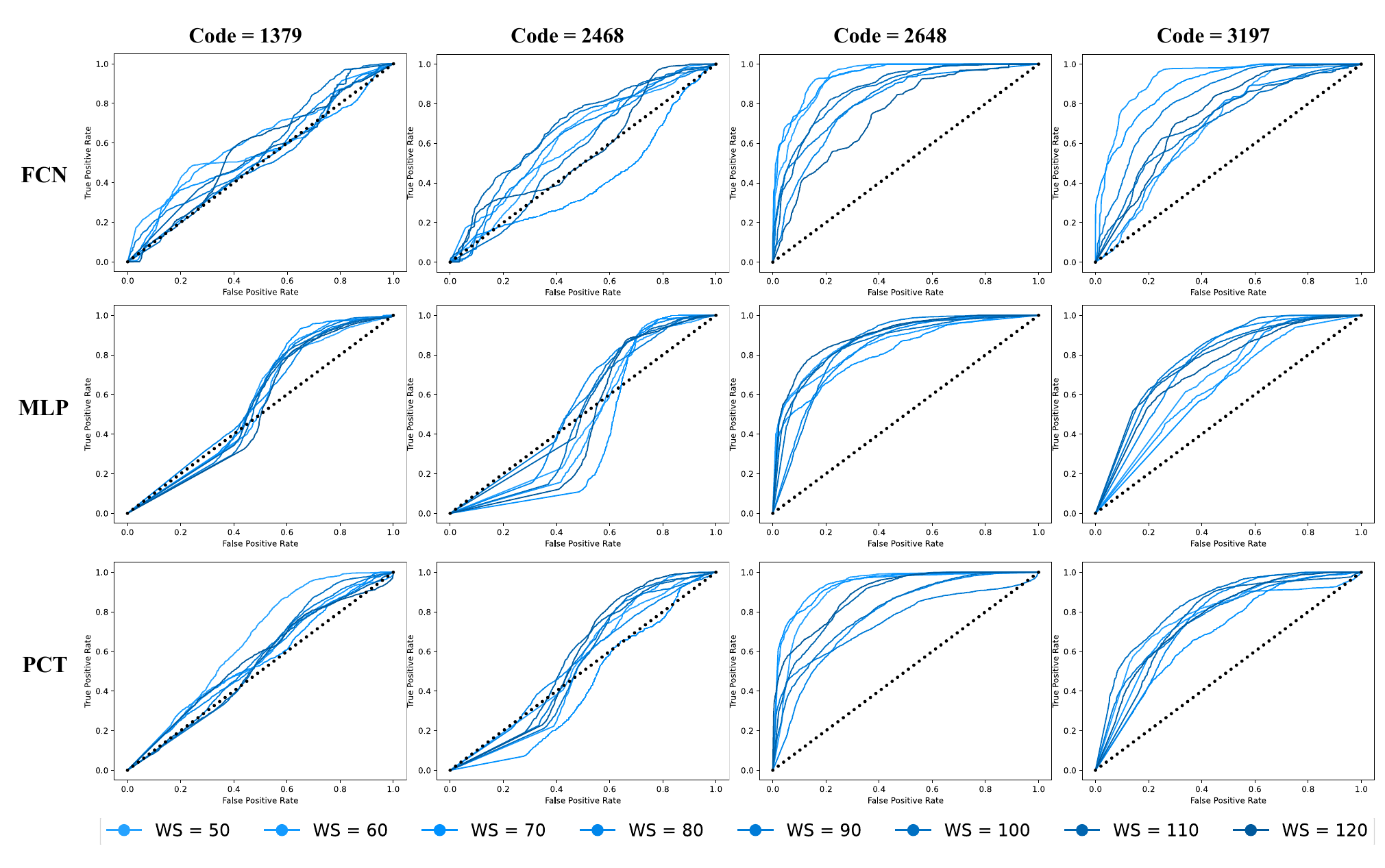}
    \caption{ROC curves for the classifiers and number combinations analyzed.}
    \label{fig:roc}
\end{figure*}

Tables~\ref{tab:classification_ACC} and \ref{tab:AUC} summarize the results acquired by running the three classifiers examined in this work with various sliding window sizes and numeric passcode combinations. We show peak test accuracy in Table~\ref{tab:classification_ACC} and area under the ROC curve in Table~\ref{tab:AUC}. Columns represent sliding windows, while rows represent classifiers and passcodes. We show the ROC curves for each of the models in Figure~\ref{fig:roc}, with columns representing passcodes and rows representing classifiers. Each curve in a plot corresponds to a sliding window size. 

As demonstrated by Table~\ref{tab:classification_ACC}, we obtain highest accuracy of 88.03\% for the combination 2648 using the PCT classifier with sliding window size of 120. We see similar high scores for other classifiers with the 2648 combination. The performance drops for smaller window sizes, with the lowest accuracy of 69.79\% obtained using an FCN. A reduction in the amount of information to learn contextual information may contribute to the drop in performance with reducing window sizes. The combination 3197 shows the next best set of accuracies, with the highest for the combination being 80.11\% using the FCN. Similar to the 2648 combination, we see a drop in accuracy with reduced sliding windows. As seen in the plots in Figure~\ref{fig:roc}, the FCN classifier shows higher overall area under the curve (AUC) for the 3197 combination compared to the MLP and PCT classifiers. PCT generally shows higher AUC for 2648. For lower window sizes, the FCN shows higher AUCs for both codes, and the MLP shows higher AUC for higher window sizes, potentially due to its simpler architecture. 

We observe that 2468 and 1379 show results close to chance. A possible reason is that given the order of entry, i.e., 2648, 2468, and 1379, users may be getting acclimatized to the environment. With the 3197 combination, unlike the other 3, the user moves from right to left. During this motion, users not familiar with VR may need re-acclimatization time, though further data collection and analysis is necessary. 

\setlength{\tabcolsep}{3pt}
\begin{table}[htb]
\caption{Classification Accuracies}
  \label{tab:classification_ACC}
  \scriptsize%
  \centering%
  \begin{tabu}{c || cccccccc}
  	\toprule
       WS & 50 & 60 & 70  & 80 & 90 & 100 & 110 & 120 \\ 
      	\hline \hline 
        MLP 1379  & 0.5234 & 0.5396 & 0.5472 & 0.5436 & 0.5427 & 0.5680 & 0.5444 & 0.5647 \\
        FCN 1379  & 0.5166 & 0.5907 & 0.5349 & 0.5275 & 0.5679 & 0.6320 & 0.6363 & 0.5453 \\
        PCT 1379  & 0.5473 & 0.5360 & 0.5291 & 0.5319 & 0.5640 & 0.5918 & 0.5925 & 0.5490 \\
        \hline 
        MLP 3197  & 0.6953 & 0.7266 & 0.7216 & 0.7144 & 0.6802 & 0.6199 & 0.5972 & 0.6054 \\
        FCN 3197  & 0.6269 & 0.7217 & 0.6766 & 0.6671 & 0.6834 & 0.6325 & 0.8011 & 0.7780 \\
        PCT 3197  & 0.6785 & 0.7053 & 0.7530 & 0.7061 & 0.7029 & 0.7438 & 0.7082 & 0.6466 \\
        \hline 
        MLP 2468  & 0.5170 & 0.5327 & 0.5062 & 0.5091 & 0.4816 & 0.4684 & 0.4586 & 0.4232 \\
        FCN 2468  & 0.6185 & 0.6497 & 0.5683 & 0.6872 & 0.6753 & 0.6586 & 0.6166 & 0.5768 \\
        PCT 2468  & 0.5623 & 0.5306 & 0.5190 & 0.5569 & 0.5269 & 0.5269 & 0.5102 & 0.5152 \\
        \hline 
        MLP 2648  & 0.7387 & 0.7149 & 0.7869 & 0.7841 & 0.8137 & 0.8130 & 0.8162 & 0.8001 \\
        FCN 2648  & 0.6979 & 0.7348 & 0.7250 & 0.7408 & 0.7126 & 0.7998 & 0.8103 & 0.8362 \\
        PCT 2648  & 0.7011 & 0.7522 & 0.7340 & 0.7762 & 0.8001 & 0.8397 & 0.8603 & 0.8803 \\
        \bottomrule
   \end{tabu}
\end{table}

% \setlength{\tabcolsep}{3pt}
% \begin{table}[htb]
% \caption{Precision Scores}
%   \label{tab:Precision}
%   \scriptsize%
%   \centering%
%   \begin{tabu}{c || cccccccc}
%   	\toprule
%        WS & 50 & 60 & 70  & 80 & 90 & 100 & 110 & 120 \\ 
%       	\hline \hline 
%         MLP 1379  & 0.4590 & 0.4685 & 0.4735 & 0.4627 & 0.4615 & 0.4739 & 0.4465 & 0.4536 \\
%         FCN 1379  & 0.4461 & 0.5461 & 0.4263 & 0.4447 & 0.4432 & 0.5457 & 0.5487 & 0.4086 \\
%         PCT 1379  & 0.4720 & 0.4516 & 0.4549 & 0.4034 & 0.4678 & 0.4909 & 0.4697 & 0.4156 \\
%         \hline 
%         MLP 3197  & 0.6264 & 0.6715 & 0.6624 & 0.6418 & 0.5868 & 0.5044 & 0.4925 & 0.4402 \\
%         FCN 3197  & 0.5891 & 0.7411 & 0.6311 & 0.6438 & 0.5971 & 0.5292 & 0.8475 & 0.8775 \\
%         PCT 3197  & 0.6244 & 0.6865 & 0.7365 & 0.6352 & 0.6209 & 0.6641 & 0.5891 & 0.5042 \\
%         \hline 
%         MLP 2468  & 0.3736 & 0.4018 & 0.3664 & 0.3676 & 0.3371 & 0.3162 & 0.2958 & 0.2391 \\
%         FCN 2468  & 0.5051 & 0.5461 & 0.4118 & 0.5938 & 0.5295 & 0.3390 & 0.3742 & 0.2055 \\
%         PCT 2468  & 0.4217 & 0.3947 & 0.3794 & 0.3990 & 0.3328 & 0.3442 & 0.3190 & 0.1409 \\
%         \hline 
%         MLP 2648  & 0.7197 & 0.7059 & 0.8164 & 0.8519 & 0.8651 & 0.8604 & 0.8861 & 0.8395 \\
%         FCN 2648  & 0.7446 & 0.8649 & 0.8981 & 0.8460 & 0.6779 & 0.8449 & 0.9615 & 0.9289 \\
%         PCT 2648  & 0.6980 & 0.8371 & 0.7590 & 0.7984 & 0.9060 & 0.7977 & 0.9024 & 0.8467 \\
%         \bottomrule
%    \end{tabu}
% \end{table}

% \setlength{\tabcolsep}{3pt}
% \begin{table}[htb]
% \caption{Recall Scores}
%   \label{tab:Recall}
%   \scriptsize%
%   \centering%
%   \begin{tabu}{c || cccccccc}
%   	\toprule
%        WS & 50 & 60 & 70  & 80 & 90 & 100 & 110 & 120 \\ 
%       	\hline \hline 
%         MLP 1379  & 0.5028 & 0.5402 & 0.6107 & 0.5795 & 0.6907 & 0.6986 & 0.6984 & 0.7604 \\
%         FCN 1379  & 0.4723 & 0.2740 & 0.2821 & 0.5307 & 0.2097 & 0.4934 & 0.3832 & 0.4623 \\
%         PCT 1379  & 0.3672 & 0.3786 & 0.5609 & 0.2533 & 0.4597 & 0.4774 & 0.3415 & 0.4853 \\
%         \hline 
%         MLP 3197  & 0.6901 & 0.6737 & 0.6567 & 0.6611 & 0.6430 & 0.7015 & 0.5005 & 0.7188 \\
%         FCN 3197  & 0.3849 & 0.5109 & 0.5109 & 0.3914 & 0.6100 & 0.4055 & 0.5684 & 0.4397 \\
%         PCT 3197  & 0.5989 & 0.5395 & 0.6206 & 0.6374 & 0.6363 & 0.6760 & 0.7156 & 0.6205 \\
%         \hline 
%         MLP 2468  & 0.3822 & 0.5071 & 0.4912 & 0.5562 & 0.5677 & 0.5903 & 0.6248 & 0.5646 \\
%         FCN 2468  & 0.2875 & 0.3824 & 0.4377 & 0.3400 & 0.2681 & 0.0872 & 0.4747 & 0.2227 \\
%         PCT 2468  & 0.3789 & 0.4758 & 0.5084 & 0.5218 & 0.4113 & 0.5529 & 0.6041 & 0.1720 \\
%         \hline 
%         MLP 2648  & 0.5531 & 0.4533 & 0.5663 & 0.5054 & 0.5719 & 0.5517 & 0.5141 & 0.4522 \\
%         FCN 2648  & 0.3567 & 0.3744 & 0.3112 & 0.3678 & 0.3849 & 0.5199 & 0.4485 & 0.5199 \\
%         PCT 2648  & 0.4272 & 0.4477 & 0.4389 & 0.5301 & 0.4966 & 0.7222 & 0.6511 & 0.7571 \\
%         \bottomrule
%    \end{tabu}
% \end{table}

\setlength{\tabcolsep}{3pt}
\begin{table}[htb]
\caption{Area under the ROC curve}
  \label{tab:AUC}
  \scriptsize%
  \centering%
  \begin{tabu}{c || cccccccc}
  	\toprule
       WS & 50 & 60 & 70  & 80 & 90 & 100 & 110 & 120 \\ 
      	\hline \hline 
        MLP 1379  & 0.5243 & 0.5397 & 0.5556 & 0.5487 & 0.5653 & 0.5896 & 0.5721 & 0.6034 \\
        FCN 1379  & 0.5114 & 0.5514 & 0.5015 & 0.5279 & 0.5133 & 0.6091 & 0.5907 & 0.5289 \\
        PCT 1379  & 0.5262 & 0.5164 & 0.5333 & 0.4924 & 0.5481 & 0.5729 & 0.5473 & 0.5364 \\
        \hline 
        MLP 3197  & 0.6946 & 0.7190 & 0.7118 & 0.7058 & 0.6738 & 0.6352 & 0.5972 & 0.6054 \\
        FCN 3197  & 0.5943 & 0.6916 & 0.6514 & 0.6225 & 0.6707 & 0.5899 & 0.7538 & 0.7028 \\
        PCT 3197  & 0.6678 & 0.6817 & 0.7329 & 0.6950 & 0.6914 & 0.7311 & 0.7097 & 0.6408 \\
        \hline 
        MLP 2468  & 0.4916 & 0.5276 & 0.5030 & 0.5199 & 0.5030 & 0.5013 & 0.5077 & 0.4693 \\
        FCN 2468  & 0.5560 & 0.5961 & 0.5403 & 0.6072 & 0.5741 & 0.5045 & 0.5747 & 0.4614 \\
        PCT 2468  & 0.5277 & 0.5196 & 0.5167 & 0.5488 & 0.4982 & 0.5339 & 0.5380 & 0.4033 \\
        \hline 
        MLP 2648  & 0.7064 & 0.6669 & 0.7442 & 0.7268 & 0.7608 & 0.7520 & 0.7405 & 0.7062 \\
        FCN 2648  & 0.6385 & 0.6687 & 0.6448 & 0.6642 & 0.6410 & 0.7345 & 0.7198 & 0.7508 \\
        PCT 2648  & 0.6534 & 0.6963 & 0.6768 & 0.7256 & 0.7338 & 0.8123 & 0.8080 & 0.8470 \\
        \bottomrule
   \end{tabu}
\end{table}

\section{Conclusion}
We demonstrate results of using deep networks to classify familiarity of a user in a VR environment using their movement patterns as they interact with a virtual keypad on a door. The task of opening a door after entering a combination is familiar to all users from having opened similar doors in the real world. We obtain highest accuracies upwards of 80\% for combinations 2648 and next highest for 3197. Results indicate potential for the use of VR movements to detect familiarity, though the effect of time on acclimatization may need to be considered, i.e., conducting familiarity detection early is critical. As part of ongoing work, we are working on expanding our data collection to include a diverse array of users and VR tasks of varying complexity. With a broader dataset, we intend to expand prior familiarity from a binary label to a 5-point Likert scale. We are also interested in using eye and head movements to gauge whether distractions from the novelty of the environment play a role in familiarity detection, e.g., if a novice user spends time examining a new environment or is taken aback. As part of future work, we are interested in investigating whether on-demand training to familiarize novice users of VR is more effective than adapting the VR environment or controller to have a reduced interaction scope. The reduced interaction scope may be actualized by reducing the amount of movement in the VR environment when the person moves the analog stick, or adding pop-up hints when performing an interaction in the environment such as opening a door. Our current familiarity detection task in VR uses a door unlock panel which requires interactions with the dominant hand. Future studies on more complex tasks, or even unfamiliar tasks, that require both hands, or controllers, is of interest as most activities in VR involve a full body experience.

\bibliographystyle{IEEEtran}
\bibliography{biblio.bib}

\end{document}